# ON THE PHILOSOPHY OF BITCOIN/BLOCKCHAIN TECHNOLOGY: IS IT A CHAOTIC, COMPLEX SYSTEM?

**Accepted (peer-reviewed) Version**[1]

RENATO P. DOS SANTOS

**Abstract:** The Philosophy of Blockchain Technology is concerned, among other things, with its ontology, how might it be characterised, how is it being created, implemented, and adopted, how does it operate in the world, and how does it evolve over time. Here, we concentrate on whether Bitcoin/blockchain can be considered a complex system and, if so, whether a chaotic one. Beyond mere academic curiosity, a positive response would raise concerns about the likelihood of Bitcoin/blockchain entering a 2010-Flash-Crash-type of chaotic regime, with catastrophic consequences for financial systems based on it. This paper starts by enhancing the relevant details of Bitcoin/blockchain ecosystem formed by the blockchain itself, bitcoin end-users (payers and payees), capital gain seekers, miners, full nodes maintainers, and developers, and their interactions. Secondly, the Information Theory of Complex Systems is briefly discussed for later use. Finally, blockchain is investigated with the help of Crutchfield's Statistical Complexity measure. The low non-null statistical complexity value obtained suggests that blockchain may be considered algorithmically complicated, but hardly a complex system and unlikely to enter a chaotic regime.

Keywords: Philosophy of Blockchain Technology, Bitcoin, complex systems, Information Theory, chaotic systems

[1] This is the peer reviewed version of the following article: On the Philosophy of Bitcoin/Blockchain Technology: Is It a Chaotic, Complex System?, which has been published in final form at Metaphilosophy 48(5):620–633, Oct. 2017. This article may be used for non-commercial purposes in accordance with Wiley Terms and Conditions for Self-Archiving.

## Introduction

Since 1990's dot.com bubble, advances in information technology provided context for an unprecedented emergence (and sinking) of various *digital currencies*, including e-gold, e-Bullion, Liberty Reserve, Pecunix, and many others, their main attractive being allowing for instantaneous transactions and borderless transfer-of-ownership. This concept also includes 'fictional' *virtual currencies* only used in the context of particular virtual communities, such as *World of Warcraft*'s in-game gold and *Second Life*'s *Linden Dollars*, or frequent-flyer and credit card's loyalty programs points. More recently, we are witnessing a surge of *cryptocurrencies*, digital currencies that use cryptographic protocols and algorithms to secure their transactions and the creation of additional units, being Bitcoin[2], created by someone under the mysterious Japanese pseudonym of Satoshi Nakamoto (2008), by far the best-known one.

More than just yet another new alternative unit of account, Bitcoin real uniqueness lays in providing a system for electronic transactions that does not rely on trust (Nakamoto 2008). In place of a centralised ledger, secured by a third-party clearer, be it a bank, a credit card company, or even a quasi bank like PayPal, this feature provides free access to a verified copy of the chain of blocks, the '*blockchain*', representing all transactions made over time, but not including any personal identifiers that can be hacked.

---

[2] We will follow here the common practice of distinguishing 'Bitcoin' (singular with an upper case letter B) to label the protocol, software, and community, and 'bitcoins' (with a lower case b) to label units of the currency, which is represented as BTC, or a capital letter B with two vertical lines "going through" it, was created by Satoshi Nakamoto himself.

Even more remarkably, besides supporting this “new form of money”, by combining BitTorrent peer-to-peer file sharing with public key cryptography (Swan 2015, 2), blockchain technology can also be used to record virtually everything that can be expressed in code: birth and death certificates, marriage licenses, deeds and titles of ownership, financial accounts, medical records, votes, the provenance of food, and so on (Swan 2015, 9–28). Right now, a blockchain-based Governance 2.0 initiative is using blockchain to allow refugees in Europe to create digital identifications that can be used to cryptographically prove their existence, which their families are, and to receive and spend money without bank accounts (Rowley 2015). This last point naturally raises issues such as cryptography, cybersecurity, and money laundering, which will not be considered here as they have been already largely discussed by many authors. Going beyond currency, which can be understood as a ‘Blockchain 1.0’ level, and contracts, as a ‘Blockchain 2.0’ one, Swan foresees a ‘*Blockchain 3.0*’stage where blockchain technology is applied to other areas, such as health, science, literacy, culture, and art (Swan 2015, 55–78).

On the other hand, we are already witnessing a skyrocketing number of smart ‘things’ – physical or virtual objects embedded with electronics, software, sensors, and connectivity through the Internet (Tapscott and Tapscott 2016). In addition, this ‘Internet of Things’ (IoT) is beginning to evolve in gathering knowledge from their interactions with other ‘things’ and humans linked to the Internet, ultimately being referred to as the ‘Internet of Everything’ (O’Leary 2013). Accordingly, Tapscott and Tapscott argue that the Internet of Everything needs a Ledger of Everything and business, commerce, and economy need a Digital Reckoning (2016), resembling Swan’s vision of a “connected world of computing” relying on blockchain cryptography (Swan 2015, 49).

All these possibilities suggest a new world of which we need to make sense, a task in which Philosophy could help us by providing a conceptual framework about how to do this sense-making.

Before proceeding with the investigation whether blockchains can be considered complex systems, the following two sections are intended to guide the non-familiarized reader through the relevant details of this blockchain technology and the most important points of the Information Theory that will be needed in the subsequent analysis.

**Bitcoin/Blockchain technology**

As mentioned above, groups of bitcoin transaction records are represented within a block, whose valid connection to the remaining blockchain must be verified and approved.

Copies of the proposed transactions are broadcast to a vast, decentralised peer-to-peer mesh network of nodes with "flat" topology architecture on top of the Internet, running the Bitcoin client software that takes the job of validating the transactions. There are no 'special' nodes in this network, and they all share the burden of providing network services.

Each block contains at its head, among other things,

- the 64-hex cryptographic hash[3] of the previous (parent) block in the chain,
- the 64-hex hash that summarises the Merkle tree[4] of the transactions included in this block,

---

[3] A cryptographic hash is a fixed length digest that results from the application of a mathematical algorithm to data of arbitrary size that is easily made, but very hard to be reverted to the original data (NIST 2012).

- a 32-bit random number, called the *nonce*, used for the proof-of-work algorithm (Antonopoulos 2014, 187).

The validation occurs when the application of SHA-256[5] to the concatenation of these three strings results in a block header hash that is numerically smaller than the network's *difficulty target* defined by the Bitcoin protocol (Nakamoto 2008). As the result of a hash function is virtually unpredictable and irreversible, the only way to validate a block is to try repeatedly, randomly modifying the nonce value until a hash matching the specific target appears by chance (Antonopoulos 2014, 188).

If a client finds such a nonce, the so-called *proof of work*, then the client has successfully validated the block, which is then added to the top of the blockchain. At the moment of this writing, with a rising network production of 4,887,867.46 Tera ($10^{12}$) hashes per second, the difficulty is adjusted to 595,921,917,085.42 (blockchain.info n.d.) so that the valid hash must be smaller than the corresponding $4.5240046586784463 \times 10^{55}$ target[6]. The last block included in the blockchain was of

[4] A Merkle tree is a tree in which the leaves are hashes of data blocks in a large data structure, allowing efficient and secure verification of the content (Antonopoulos 2014, 164).

[5] SHA-256 is a cryptographic hash function designed by the National Security Agency (NSA) that produces hashes of 256 bits from inputs of any length less than $2^{64}$ bits (NIST 2012).

[6] The target is calculated by dividing the maximum target used by SHA-256 (which should logically correspond to 256 1 binary digits, but, because Bitcoin stores the target as a floating-point type, this is truncated to approximately $2^{224}$, which can be represented as the 64-hex hash

#468716, with the 0x000000000000000000ca45d84c525fb48ddc162bdf793a47bf2d136ad938733c 64-hex hash[7], which corresponds to 1.937388828842762×10$^{55}$ and is indeed smaller than the difficulty target. Many Bitcoin users talk about the “number of leading zero bytes in the hash,” instead of “lower than the target” – in this example, there are 18 leading zeros in the hex representation of the block hash. Whereas this is not exact, it will be convenient in the calculations ahead.

Furthermore, if the client is the first to publish the validated block into the blockchain, it is accordingly rewarded with a sum of newly-created bitcoins plus the transaction fees from all the transactions included in that block (Antonopoulos 2014, 173). This compensation works as an incentive scheme that not only aligns the actions of client users with the security of the network but also simultaneously implements the monetary supply of bitcoins (Antonopoulos 2014, 174).

Bitcoin protocol determines that the total number of bitcoins that will ever be mined is limited to 21 million and that it should be reached c. 2137. As it also establishes that the network's *difficulty* target is adjusted every 2,016 blocks, to make sure that a new block is created about every 10 minutes (Antonopoulos 2014, 2), the value of the reward, which presently amounts to 12.5 BTC plus transaction fees, is halved every 210,000 blocks or about every four years to limit the supply of bitcoins and avoid inflation.

---

0x00000000FFFF0000000000000000000000000000000000000000000000000000, being “0x” the usual prefix to flag hexadecimal numerals) by the difficulty.

[7] Bitcoin represents 256-bit hashes as strings of 64 hexadecimal symbols (Antonopoulos 2014, 64).

The smaller the hash must be, the rarer it is, and the harder to find a nonce that leads to it. Quintillions of random nonce values are tested before a compliant one is found. Due to the effort needed, the process of finding proper nonces is compared to extracting precious metals from the earth, and it is thus called *mining*, whilst the participant is referred to as a *miner* (Antonopoulos 2014, 174).

Mining is the primary activity by which blockchain operates, and it is related to information theory and complex systems, as shown later.

As said before, the hash of the content of one block is unpredictable. However, as the hash of one block enters in the determination of the hash of the next block, it also is not entirely independent of the content of the previous one.

Generally, a small change to the parent hash causes the children hash so much change as to appear uncorrelated with the old one; yet, collisions – the existence of two or more different parent blocks $x_1, x_2, x_3, \cdots$ that lead to the same hash $H(x_1) = H(x_2) = H(x_3) = \cdots$ and, consequently, to the same child block hash – are not impossible. If the hashes are evenly produced, the probability of a collision among $k$ distinct pieces of data to 256-bit hashes comes from the mathematics behind the *birthday problem* (Rouse Ball 1960, 45) as given by $k^2/2^{257}$ for large $k$. For the $k = 431{,}616$ blocks already mined at the time of this writing, the minuscule probability of a collision among them is $8.0 \times 10^{-67}$; a collision is, therefore, expected only after $2^{256/2} = 3.4 \times 10^{38}$ blocks have been mined, which being a new block mined about every 10 minutes, can take about $6.5 \times 10^{33}$ years to happen[8].

---

[8] Another way of grasping this unlikelihood is to consider that there are $2^{256}$ different 256-bit hashes (approximately $10^{77}$ in decimal), whilst the entire visible universe is estimated to contain $10^{80}$ atoms (Antonopoulos 2014, 64).

Antonopoulos argues that Nakamoto's main invention was making all the properties of Bitcoin, including currency, transactions, payments, and the security model, derive from its decentralised mechanism for *emergent* consensus. A consensus that is not achieved explicitly – by election or decision from some central authority –, nor by any inherent complexity or trust in any single node. Instead, it is an artefact of the asynchronous interaction of a resilient network of thousands of uncomplicated, independent nodes, all following straightforward, algorithmic rules to accomplish a myriad of financial processes (Antonopoulos 2014, 177).

Notwithstanding, some authors assume that the whole blockchain code is characterised by a high degree of complexity, apparently confusing the high complication (of the code) with an eventual complexity of the resulting blockchain and Bitcoin ecosystem. Consequently, blockchain seems worth an analysis through Complexity Theory to clarify this issue.

## Information theory of complex systems

Complex systems are often quite loosely defined as those constituted by many different components in strong interaction and showing some self-organization and emergent structured collective behaviour. Nevertheless, "complexity" has been so much used without qualification by so many authors, both scientific and non-scientific, that the term has, unfortunately, almost lost its meaning (Feldman and Crutchfield 1998).

There is no concise definition of a complex system, nor a consensus set of definite properties associated with the concept of a complex system, being non-linearity, chaotic behaviour, presence of feedback, robustness, emergent order, and hierarchical organization, among others, usually mentioned in the literature as necessary, but not sufficient (Ladyman, et al. 2013). Consequently, instead of pursuing a definition of a complex system, we follow Ladyman, et al. in opting for characterising a system by some measure of its complexity. In the following, we will discuss a few of the many such measures available.

It is widely known that, for many systems that are far from thermodynamic equilibrium, near phase transitions, several thermodynamic quantities, such as specific heat, compressibility and magnetic susceptibility, present singular behaviour in the critical region with asymptotic differences characterised by critical exponents that define ‘power laws.' This observation has encouraged physicists to equate power laws as such with complexity and claim that some system is complex because it exhibits a power law distribution of event sizes (Shalizi 2006, 61).

A pioneer application of this concept was the scaling behaviour of ‘fat tails’ observed in the dynamics of the Standard & Poor’s 500 index (Mantegna and Stanley 1995). More recently, a power-law correlation with an exponent between 0.8 and 1.1 was observed on Google search queries for Dow Jones Industrial Average (DJIA) component stocks (Kristoufek 2015). Likewise, a fat tail with exponents around 2 was found in the distribution of fluctuation in the exchange rates in international trade in foreign exchange market (FOREX) (Chakraborty, et al. 2016) and in the bitcoin price volatility (Easwaran, et al. 2015).

Despite its popularity, however, no one has ever demonstrated any relation between power laws and any formal complexity measure. Besides, although it is true that one does not find any power laws in equilibrium statistical mechanics when the system is not complex by usual standards, it has been known for half a century that there are many ways of generating power laws without complexity (Shalizi 2006, 61). Worse still, Shalizi shows that most of the things claimed to be power laws are actually other kinds of heavy-tailed distribution, mistakenly taken as such due to inadequate statistical estimations of the adherence of the data to a power law, apparently forgetting that *any* smooth curve looks like a straight line, if attention is confined to a sufficiently small region, which, for some non-power-law distributions, can extend over multiple orders of magnitude (2006, 61-65).

Some authors assume that complicated systems are strong candidates to complex systems, or, at least, that complex systems must be complicated. Complication and complexity are quite distinct features, however: for example, an atomic clock is a very complicated device, yet a highly predictable one and, therefore, not complex at all; conversely, a planar double pendulum is a mechanical system that exhibits complex, chaotic behaviour (Stachowiak and Okada 2006), despite its simple construction. On the other hand, systems which are strongly ordered have only a small range of allowable behaviour and cannot be complex, whereas those that are strongly disordered have independent, weakly interacting parts and are not complex either (Shalizi 2006, 54).

This relation between complexity and order (as measured by entropy) is sketched in Fig. 1. At the lower extreme, ordered systems such as a perfect crystal have low entropy and low complexity. At the higher extreme, completely random systems such as a fair coin flip have high entropy yet low complexity. Therefore, the one point of agreement in the complex-systems community is that any satisfactory measure of complexity should be minimal both for completely random and ordered systems, as they admit of concise descriptions, and assign the highest value of complexity to systems which are neither completely random nor completely ordered, lying somewhere in the middle (Ladyman, et al. 2013).

Figure 1 – The intuitive relation of complexity and entropy.

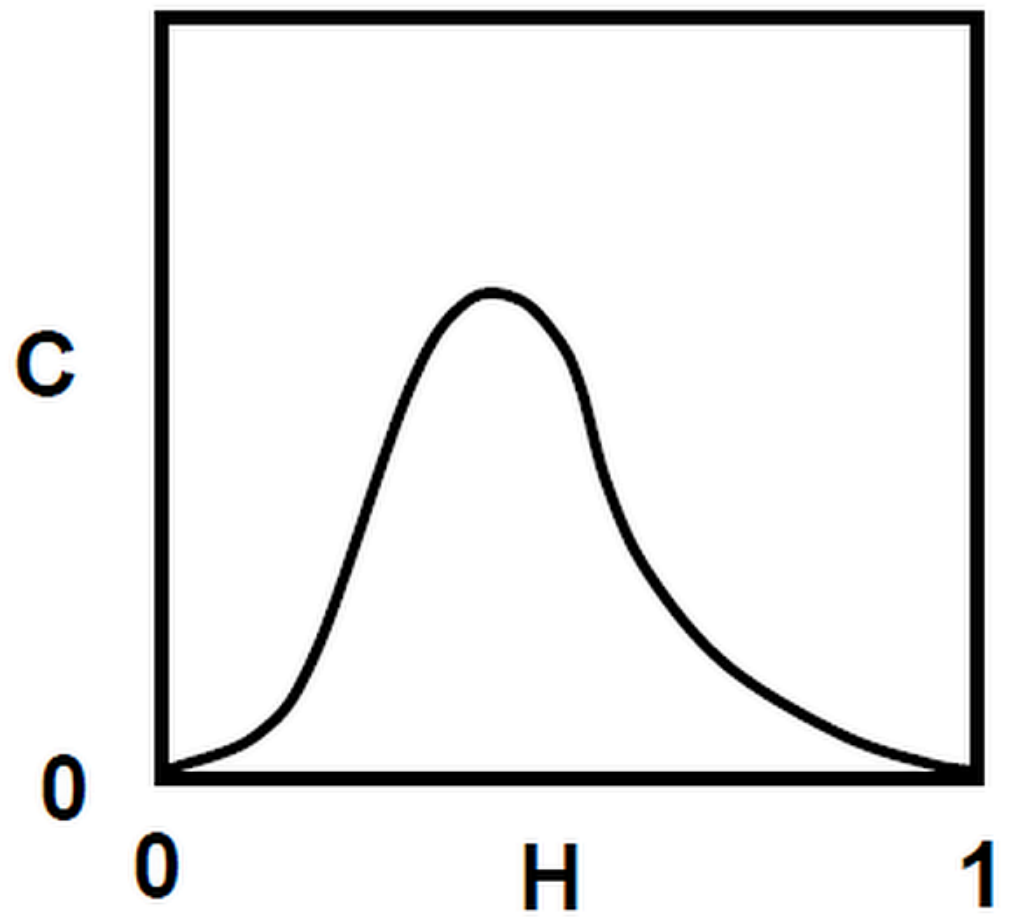

Source: (Crutchfield and Young, 1990)

Ladyman, et al. analysed various measures of complexity available in the scientific literature and showed that many of them are not computable whilst other ones are not appropriate measures of the complexity of a physical system as they are ultimately measures of randomness, rather than of complexity and give maximal values to completely random data processes (Ladyman, et al. 2013). Those authors concluded that the measure that best captures the qualitative notion of complexity of systems is that of the Statistical Complexity, introduced by James Crutchfield and Karl Young (1989).

Accordingly, here, we will make use of this alternative approach to complexity, as it is more suitable for physical systems (Crutchfield 2011). It adapts and extends ideas from the theory of discrete computation to inferring complexity in dynamical systems, namely through the notion of a 'machine' – a device for encoding the structures in discrete processes (Crutchfield 1994). Once we have reconstructed the machine, we can say that we understand the structure of the process and talk about the structure of the original process regarding the complexity of the reconstructed machine (Crutchfield 1994).

The input to computation is given by the system's initial physical configuration; performing the computation corresponds to the temporal sequence of changes in the system's internal state; and the result of the computation is read off finally in the state to which the system relaxed (Crutchfield 1994). These computational models reconstructed from observations of a given process are referred as $\epsilon$-machines, in order to emphasise their dependence on the limited accuracy $\epsilon$ of the measuring instrument (Crutchfield and Young 1989).

Statistical complexity then measures the structure of the minimal machine reconstructed from observations of a given process regarding the machine's size (Crutchfield 1994). Therefore, statistical complexity describes not only a system's statistical properties but also how it stores and processes information and intends to understand its randomness and organisation from the available, indirect measurements that an instrument provides (Crutchfield 2011).

Three optimality theorems ensure us that the $\epsilon$-machine representation of a process captures all the process's properties: by being its optimal predictor, its minimal representation compared with all other optimal predictors, and being any other minimal optimal predictor equivalent to the $\epsilon$-machine (Crutchfield 2011).

The system may be characterised by a set of observable random variables $X_i$ , having at each instant $t$ a past $X^- = \cdots, X_{t-2}, X_{t-1}$ and a future $X^+ = X_{t+1}, X_{t+2}, \cdots$. There are possibly various, different histories of measurements $x_i^-$, from which a prediction, that is, a conditional probability distribution $P(X^+|x_i^-)$ to the future observations $X^+$ can be defined, with which anyone can predict everything that is predictable about the system. Two histories $x_1^-$ and $x_2^-$ are said to be equivalent if $P(X^+|x_1^-) = P(X^+|x_2^-)$; classes $\sigma$ of equivalent histories may be then constructed, and one may stop distinguishing histories that make the same predictions. These equivalence classes $\sigma$ are called the 'causal states' of the process, as they retain themselves the relevant information for predicting the future (Crutchfield 2011).

Given this framework, the statistical complexity $C_\mu$ can be calculated as

$$C_\mu = -\sum_{\sigma \in S} P(\sigma) \log_2 P(\sigma)$$

where $S$ is the set of all the causal states and $P(\sigma)$ is the probability distribution over them (Crutchfield 2011).

It may be enlightening here to apply this formula to the previous two examples, following Crutchfield calculations (2011). A crystal has one single state (its unchanging characteristic spatial arrangement of its composing atoms extending in all directions) and, consequently, there is only one class $\sigma$ of equivalent histories, $P(\sigma) = 1$, and the statistical complexity $C_\mu = 0$, as desired. For the other example, at any given instant the future observations $X^+$ of the fair-coin flip is unpredictable; all different histories of measurements $x_i^-$ are equivalent in their inability to predict it, there is only one class $\sigma$, $P(\sigma) = 1$, and again $C_\mu = 0$.

Introducing a third example, consider a period-2 process such as a perfectly periodical pendulum clock, which after being put in motion, exhibits two recurrent causal states, say *tick* and *tock*; now, there are two distinct histories leading to each of these states, $P(\sigma_{tick}) = P(\sigma_{tack}) = 1/2$, and, in sharp contrast to the first two examples, it shows itself quite surprisingly as a structurally complex process with $C_\mu = 1$.

These results show that the Statistical Complexity of Crutchfield and Young does satisfy the intuitive requirements for a measure of complexity discussed above (Fig. 1) and support our methodological choice of it for this work.

We now proceed to measure the Statistical Complexity of the stream of blocks provided by the blockchain to infer whether the blockchain ecosystem can be considered a complex system.

**On the complexity of blockchain**

Following Crutchfield and Young, blockchain can be seen as an infinite-string-production machine that oscillates between two states about every 10 minutes:

1. Initial state: a new block was just incorporated into the blockchain, and the machine starts mining a new block that includes most of the pending transactions collected from around the world into the *transaction pool*. Hashes are generated and tested against the network's *difficulty target*.
2. A nonce that results in a hash smaller than the target is found, the validated block is broadcast to the P2P network for inclusion into the blockchain and, even if a blockchain *fork* (Antonopoulos 2014, 199) happens, the global Bitcoin network ultimately converges to a new consistent initial state.

As collisions – the existence of two or more different parent blocks that lead to the same hash and the same child block hash – are highly improbable, the classes $\sigma$ can be taken as including only one possible block each.

In the example above, there are 18 leading zeros in the hex representation of the target. Now, there are $16^{64-18} \cong 2.45 \times 10^{55}$ different 64-hex strings with 18 leading zeros and this allows the probability of finding a hash that would fulfil the present difficulty target to be estimated as $P(18) = 16^{64-18}/16^{64} \cong 2.1 \times 10^{-22}$.

Therefore, the probability of the 18 leading-zeros-hash causal state $\sigma_2$ is $P(\sigma_2) \cong 2.1 \times 10^{-22}$, the probability of the initial state $\sigma_1$ is $P(\sigma_1) \cong 1 - 2.1 \times 10^{-22}$, and the statistical complexity $C_\mu$ can be calculated[9] as

$$C_\mu = -\begin{bmatrix} (1 - 2.1 \times 10^{-22}) \log_2(1 - 2.1 \times 10^{-22}) \\ +(2.1 \times 10^{-22}) \log_2(2.1 \times 10^{-22}) \end{bmatrix}$$

or

$$C_\mu \cong 1.56 \times 10^{-20}$$

This extremely low statistical complexity result leads us to the conclusion that blockchain can hardly be considered complex, confirming Nakamoto's statement that "the network is robust in its unstructured simplicity" (2008). The functioning of the blockchain may be regarded as algorithmically complicated, but not complex. There is no nonlinearity apparent in the blockchain, or any hierarchical growth, or emergence, and the regularly-distributed production of highly replicated versions of the similar blocks about every 10 minutes does not imply complexity.

Differently from the periodical pendulum clock example above, where $P(\sigma_1) = P(\sigma_2) = 1/2$ and $C_\mu = 1$, the fact that $P(\sigma_1) \gg P(\sigma_2)$ here led to this very small value for the statistical complexity $C_\mu$. As blockchain can be represented by an infinite-string production machine that oscillates between two states, and the mining nodes spend a huge part of their time producing random hashes that do not make the system to transition to a new block state, its complexity could be expected to be not far from the fair-coin flip, namely zero. Notice that, even if blockchain may be seen as an entropy-generation system[10], because the next source or node of the correct nonce guess cannot be anticipated, it was already shown above that high-entropy, completely random systems have yet low complexity.

---

[9] Due to $(\sigma_2)$ being much smaller than 1, we used Shanks transformation $\log_e(1 - x) \cong -x(6 - x)/(6 - 4x)$ to increase the precision of calculation of the $(1 -$ $P\sigma 2$log2$1-P\sigma 2$ term.

It is worthwhile highlighting that, beyond mere academic curiosity, a positive response to the question whether blockchain is a complex system would raise concerns about the likelihood of blockchain spontaneously entering an emergent 2010-Flash-Crash-type of chaotic regime (Twin 2010), with catastrophic consequences for financial systems based on it.

To investigate this issue, it seems appropriate to distinguish the Bitcoin digital payment system from the blockchain distributed ledger that sustains it. In fact, according to Pilkington (2016), whereas bitcoin transferability is ensured by the blockchain technology, it is its acceptability that makes bitcoin money, and this can only be explained by pure market behaviour.

From Crutchfield and Young (1990), one learns that critical transitions to chaotic regime usually happen in the middle state region between low entropy and high entropy, where complexity attains its maximum (Figure 1). Now, from the previous result that blockchain infrastructure (including miners, full nodes maintainers, and developers) presents an extremely low statistical complexity measure, it is reasonable to assume that blockchain is also hardly expected to enter a chaotic regime.

The same cannot, unfortunately, be said of the Bitcoin ecosystem built upon blockchain, which includes not only functional bitcoin end-users (payers and payees), but also capital gain seekers who have no functional usage for the currency apart from an expectation of capital gains, hoping to buy bitcoins cheap and sell them expensive (2014), leading to eventual ‘black swan’ events such as the recent Mt. Gox bitcoin exchange bankruptcy (Takemoto & Knight 2014).

[10] Actually, coders have been discussing using blockchain as a truly, fully non-arbitrary, publicly available, cheaply distributed, random generator with timestamp (see e.g. https://github.com/billpmurphy/blockchain-random).

Even if the study of Easwaran, et al. (2015) is disregarded for being a debatable ‘power law’ one, Siddiqi (2014) shows that the bitcoin market is a complex system without a stable equilibrium in which the presence of the capital gain seekers introduces nonlinearity, which makes chaos likely to arise, be it via the logistic map if this category of users prevails over the functional ones, or via the delay logistic-Henon map if both categories matter. To limit the relevance of capital gain seekers and, therefore, reduce nonlinearity and prevent chaos, Siddiqi suggests a regulation constraining online exchanges from bitcoins into dollars and vice versa to buys and sells of real goods and services (2014).

**Conclusion**

This paper, based on the Philosophy of Blockchain Technology, discussed what blockchain is, how is it being created, how does it operate in the world, and how does it evolve over time. It concluded that the blockchain ecosystem, formed by the blockchain itself, bitcoin end-users (payers and payees), miners, full nodes maintainers, and developers, and their interactions, that keeps its distributed ledger working might be considered algorithmically complicated, but hardly a complex system and unlikely to enter a chaotic regime. Relying on the resilience and irreversibility of the blockchain, users can safely keep recording digital documents as well as issuing digital identification to refugees. According to Siddiqi (2014), however, the presence of the capital gain seekers in the bitcoin market makes chaos likely to arise, an undesirable scenario that could be prevented if some measures were taken to limit the role of capital gain seekers.

*PPGECIM - Doctoral Program in Science and Mathematics Education*

*ULBRA - Lutheran University of Brazil*

*Av. Farroupilha, 8001*

*92425-900 Canoas, RS*

*Brazil*

*renatopsantos@ulbra.edu.br*